\documentclass[showpacs,twocolumn,floatfix,aps,pre,reprint]{revtex4-1}

\usepackage{natbib}
\usepackage{tikz}
\usepackage{gensymb}

\newcommand\Rey{\mbox{Re} }

\usepackage[breaklinks=true,pdfstartview=FitH,
            CJKbookmarks=true,
            bookmarksnumbered=true,
            bookmarksopen=true,
            colorlinks=true,
            pdfborder=001,
            linkcolor=blue,
            citecolor=blue,
            urlcolor=blue
            ]{hyperref}

\begin{document}

\title{Crisis bifurcations in plane Poiseuille flow}
\author{Stefan Zammert and Bruno Eckhardt}
\affiliation{Fachbereich Physik, Philipps-Universit\"at Marburg, 
D-35032 Marburg, Germany}
\affiliation{J.M. Burgerscentrum, Delft University of Technology, 
2628 CD Delft, The Netherlands}

\begin{abstract}
Many shear flows follow a route to turbulence that has striking similarities
to bifurcation scenarios in low-dimensional dynamical systems. Among
the bifurcations that appear, crisis bifurcations are important because 
they cause global transitions  between open and closed attractors, or 
indicate drastic increases in the range of the state space that is 
covered by the dynamics.
We here study exterior and interior crisis bifurcations in
direct numerical simulations of transitional plane Poiseuille flow in
a mirror-symmetric subspace.
We trace the state space dynamics from the appearance of the first 
three-dimensional exact coherent structures 
to the transition from an attractor to a chaotic saddle in an exterior crisis. 
For intermediate Reynolds numbers, the attractor undergoes several interior crises, 
in which new states appear and intermittent behavior can be observed. 
The bifurcations contribute to increasing the complexity of the dynamics and to a more dense
coverage of state space. 
\end{abstract}

\maketitle

Numerical and experimental studies of pipe and plane Couette flow have demonstrated the significance
of exact coherent structures and their bifurcations for the transition to turbulence  \cite{Hof2004,Kreilos2012,Kawahara2012,Avila2013}. Typically, these states
appear in saddle-node bifurcations and then undergo further bifurcations. 
Initially, most of their complexity lies in the temporal dynamics, so that they are better characterized 
as chaotic rather than turbulent. With increasing Reynolds number, more temporal and spatial
degrees of freedom are activated, until the complexity of a turbulent flow is established. 
Parallel to the increase in complexity comes a growth of the parts of state space that 
participate in the chaotic and turbulent dynamics. Studies of low-dimensional dynamical
systems have revealed many routes to this increased complexity  \cite{Ott2002,Strogatz1994,Tel2008,Lai2011}.
Several of them have already been discussed in the context of high-dimensional fluid
systems, e.g. in the cases of plane Couette flow  \cite{Kreilos2012} or pipe flow  \cite{Mellibovsky2012,Avila2013}.
One contribution of the present study is to document similar phenomenology in another canonical
fluid system, plane Poiseuille flow (PPF). A second one is the demonstration of 
interior crisis and their contribution to increasing the complexity of the attractor and of the state space region covered
by it.

PPF is the pressure driven flow between two parallel plates and differs from
plane Couette flow and pipe flow because of the presence of a linear instability to 
transverse vortices, the so-called Tollmien-Schlichting modes \cite{Heisenberg1924,Lin1945,Thomas1953}.
It occurs at a critical Reynolds number 
of $5772.22$ for a streamwise wavenumber $\alpha$ of $1.02056$ (based on the center-line velocity and 
half the gap width), as determined by Orszag  \cite{Orszag1971}. 
The bifurcation is  subcritical, and reaches down to about 
$\Rey\approx 2700 $ \cite{Zahn1974,Soibelman1991}
(for different wavelength). However, several experiments and numerical simulations show that
turbulence occurs already at Reynolds numbers around $1000$  \cite{Carlson1982,Lemoult2012,Tuckerman2014},
and hence well below the onset of Tollmien-Schlichting modes. Thus, the linear instability cannot 
explain the observed turbulence at low Reynolds numbers and the situation becomes analogous
to that in plane Couette and pipe flow.

In order to determine the relevant saddle-node bifurcation in PPF we use the method of edge tracking, 
as described in  \cite{Skufca2006}, see also  \cite{Toh2003}. The method traces the time-evolution 
of initial conditions and uses bisection between an initial condition that returns to the laminar profile
and one that becomes turbulent to approximate one on the laminar-turbulent interface. In most cases the
state evolves towards a simple attractor, such as a fixed point or a simple periodic orbit.
It is then possible to continue the edge state in Reynolds number around the saddle-node
bifurcation and to identify the upper branch solution.
In recent work for plane Couette \cite{Kreilos2012} and pipe flow \cite{Avila2013} it was shown that 
the upper branch of the edge state undergoes various bifurcations resulting in a chaotic attractor.
A boundary or exterior crisis ultimately  destroys this stable attractor and creates
the observed transient turbulence with its characteristic exponentially distributed lifetimes. 
The observed phenomenology is similar to what has been described and discussed in the 
context of chaotic dynamical systems  \cite{Tel2008,Lai2011}.

We will here show that this scenario is also present in plane Poiseuille flow and
that another type of crisis bifurcation, the interior crisis, 
provides a mechanism by which the part of state space occupied by the chaotic attractor can increase. 
Furthermore, we will discuss mechanisms for the observed increase of lifetimes of 
chaotic transients  \cite{Hof2006}. 

For our numerical simulations we use the \textit{Channelfow}-code  \cite{J.F.Gibson2012}.
The Reynolds number $\Rey=U_{0}d/\nu$ for the system is based on half the distance between the 
plates $d$,
the maximum velocity of the laminar profile $U_{0}$ and the kinematic viscosity $\nu$. 
We take a coordinate system in which 
$x$ points in the streamwise, $z$ in the spanwise and $y$ in the wall-normal direction. 
With the above choices for the dimensionless units the laminar profile becomes 
$\vec{u}_{l}(y)=(U(y),0,0)$  with $U(y)=1-y^{2}$. The total flow field $\vec{u}_{t}$ can be written as the 
sum of the laminar profile and a fluctuating component,  $\vec{u}_{t}=\vec{u}_{l} + \vec{u}$.
All simulations in the paper are performed for constant mass flux and with no-slip boundary conditions at the walls.
The calculations are restricted to a computational domain 
of length $2\pi$, width $\pi$ (and height $2$)
in a subspace that is symmetric to reflections at the midplane and to spanwise reflections at the plane defined by $z=0$:
\begin{eqnarray}
s_{y}: [u,v,w](x,y,z)=[u,-v,w](x,-y,z)\\
s_{z}: [u,v,w](x,y,z)=[u,v,-w](x,y,-z)
\end{eqnarray}
As in other studies  \cite{Kreilos2012,Avila2013} the restriction to a symmetric subspace 
stabilizes the exact coherent structures.

The numerical resolution is  $N_{x}\times N_{y} \times N_{z} = 48 \times 65 \times 48$ modes.
We checked higher values of $N_{x}$ and $N_{z}$ and found no significant changes 
in the bifurcations. The exact coherent structures and the location of the bifurcations vary with domain size, but the
phenomenology is similar. The same applies to the localized coherent states and their bifurcations that appear
in wider domains. Especially, in large domains the states become localized  
\cite{Melnikov2014,Zammert2014b}. 

\begin{figure}
\centering
\includegraphics{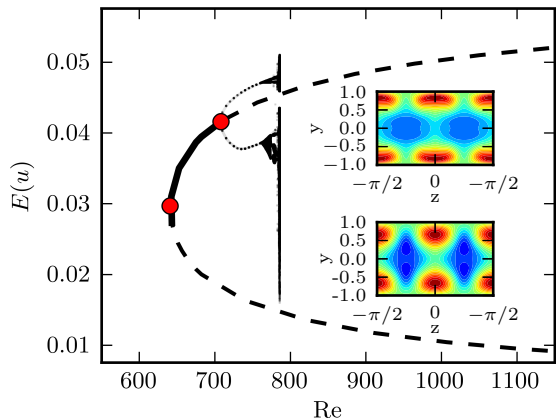}
\caption{Bifurcation diagram for the travelling wave $TW_{Eyz}$. Solid lines indicate a stable state,
dashed lines an unstable state. Chaotic and periodic states are indicated by clouds of points obtained
by plotting minima and maxima of their energy densities in the course of time.
The red dots indicate the bifurcation point of $TW_{Eyz}$ and $PO_{yz}$, respectively.
The insets show the average streamwise velocity in the spanwise-wallnormal plane for the lower and the upper branch of $TW_{Eyz}$ at $\Rey=830$.
The colors indicate low (blue) and high (red) velocity regions.
}
\label{fig_BifDiag}       
\end{figure}

Using the technique of edge tracking it is possible to identify the edge state  \cite{Skufca2006,Schneider2008} of this system.
A trajectory on the laminar-turbulent boundary quickly reaches a states of constant energy. 
Since a stationary state
is ruled out on account of the non-zero mean flow, 
the attractor in the laminar-turbulent boundary is a travelling wave. Indeed, a
Newton search  \cite{Viswanath2007} for a relative equilibrium converges to a 
traveling wave, henceforth referred to 
as $TW_{Eyz}$. The form of the state is indicated in the inset in Fig. \ref{fig_BifDiag}.
The travelling wave has the same symmetries as the mirror-symmetric travelling wave previously 
described by Nagata and Deguchi  \cite{Nagata2013a} and Gibson and Brandt  \cite{Gibson2014}.

We use a continuation method (see  e.g.  \cite{Dijkstra2013}) to follow the solutions in 
Reynolds number around the saddle-node bifurcation at $\Rey_{SN}=641$. 
A stability analysis of the travelling wave shows that in the symmetry subspace the lower branch has one unstable eigenvalue and the upper branch is stable for $641<\Rey<707$.

At $\Rey = 707$ the upper branch undergoes a Hopf bifurcation that creates a 
stable relative periodic orbit $PO_{yz}$.
This orbit undergoes a Neimark-Sacker bifurcation at $\Rey = 761.5$ that creates a stable torus. 
In further bifurcations a chaotic attractor is generated. By plotting minima and maxima of the energy
\begin{equation}
E(\vec{u})=\frac{1}{2L_{x}L_{z}} \int_{0}^{L_{z}}  \int_{-1}^{1}  \int_{0}^{L_{x}} \vec{u}^{2} dx dy dz
\end{equation}
of a trajectory on the attractor we are able to map out the bifurcation diagrams also in chaotic regions, as 
shown in Fig. \ref{fig_BifDiag}. The mapping of the chaotic region becomes feasible due to the
restriction by the shift-and-reflect symmetry that stabilizes the states.  In the full system the entire
bifurcation structure persists 
within an unstable subspace.

\begin{figure}
\centering
\includegraphics{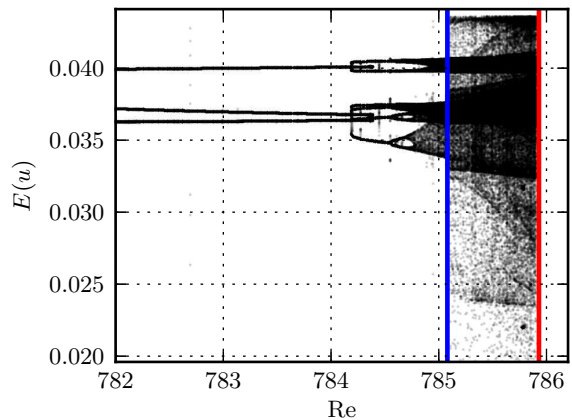}
\caption{Magnification of the attractor in Fig. 1 in order to highlight the interior crisis (blue line)
and the exterior crisis (red line). The states are visualized by 
by plotting minima of $E(u)$ along trajectories.}\label{fig_InteriorCrisis2}       
\end{figure}

The magnification of the chaotic attractor in Fig. \ref{fig_InteriorCrisis2} highlights the 
two phenomena we want to focus on here: 
Slightly above $\Rey=785$ (blue line) the size of the attractor expands and covers a larger fraction of the
interval, and slightly below $\Rey=786$ (red line) it disappears. Both changes are connected with crisis bifurcations \cite{Tel2008,Lai2011}, an interior crisis in the first case, and an exterior crisis in the second case.

\begin{figure}
\centering
\includegraphics{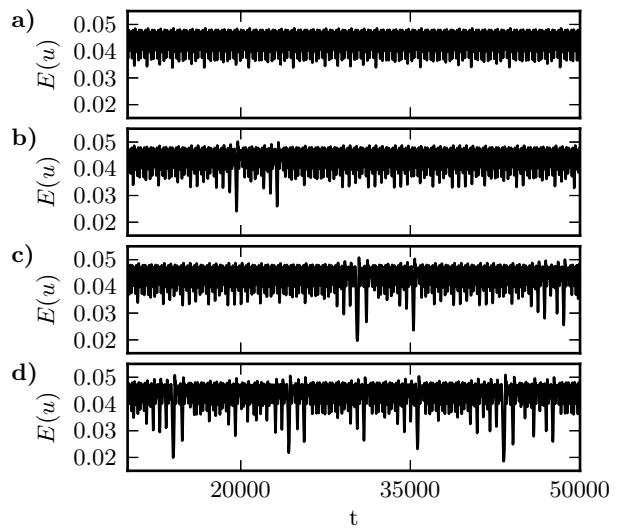}
\caption{Energy traces $E(t)$ for trajectories near the interior crisis. 
(a)  $\Rey=785.0$, slightly below the crisis. (b), (c), and (d) are for 
$\Rey=785.46$, $785.75$, and $785.9$, respectively, above the crisis.
They show the characteristic intermittent bursts.}
\label{fig_Intermittency}       
\end{figure}

Slightly above $\Rey=785$ the points on the attractor suddenly spread over a wider region, 
covering the area in state space with an energy $
E$ between 0.023 and 0.044. However, these parts of the state space are only visited occasionally,
so that the points are less dense than in other parts.
The reason for the sudden enlargement is a so-called interior crisis bifurcation  \cite{Grebogi1982,Grebogi1983},
where a new state appears and new links to the attractor form.
The appearance of the new states can be seen in the time series in  Fig. \ref{fig_Intermittency}. 
Just before the crisis, the range of the trajectories is limited to the interval 
$\lbrack 0.033,\ldots,0.042 \rbrack$. 
Slightly above the crisis, excursions to lower values occur, with their number increasing with \Rey.

\begin{figure}
\centering
\includegraphics{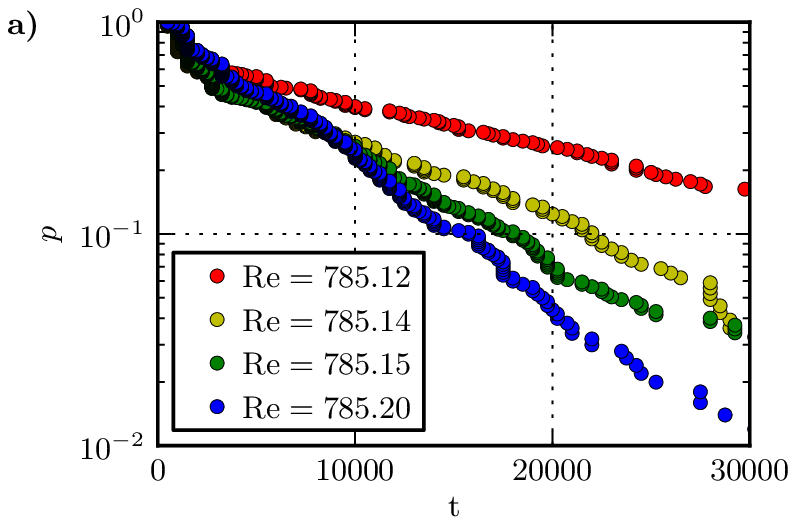}
\includegraphics{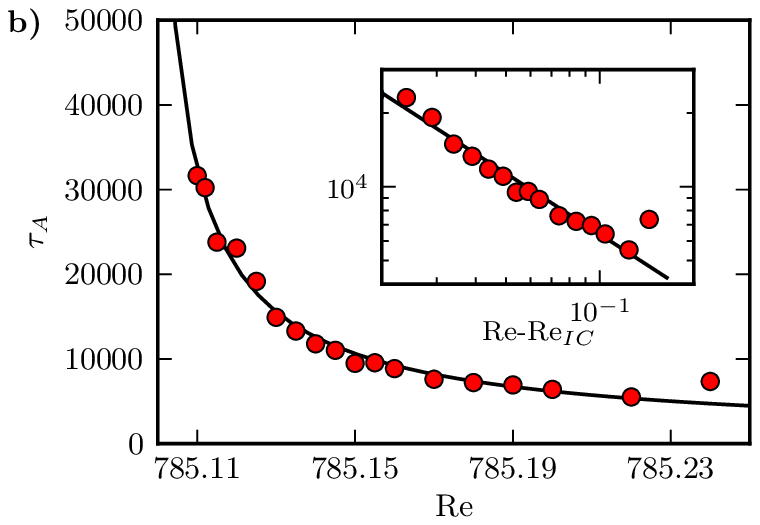}
\caption{Intermittency near the interior crisis. (a) Probability to stay $t$ time units in the pre-crisis
phase A. The times are exponentially distributed and the characteristic timescale $\tau_{A}$ 
increases with decreasing distance to $\Rey_{IC}$.
(b) Variation of characteristic times with Reynolds number. The continuous line shows an 
algebraic fit to equation (\ref{eqn_IC}) with $\gamma=0.8$. The inset shows the  data 
on a doubly logarithmic scale where on the abscissa the distance to the Reynolds number of the interior crisis is used.
 }\label{fig_InteriorCrisis}     
\end{figure}

The type of transition can be determined from the distribution of times spent in the different regions.
The state space region covered by the attractor before the crisis, referred to as phase $A$, 
contains trajectories that never drop below a threshold in energy, here
taken to be $E_{t}=0.031$. If the trajectory drops below $E_{t}$, trajectories enter a different
phase $B$, occupying a different region in state space. An indicator for phase $B$
are the repeated excursions to values below $E_{t}$. Accordingly, if no excursions are noted
for more 250 time units, we conclude that the system has returned to phase $A$.
With this prescription one can determine the distribution of times in phase $A$ as shown in 
Fig. \ref{fig_InteriorCrisis}a. The plot contains data from trajectories with a total length of 
$5\cdot10^{6}$ time units. The data are shown semi-logarithmically,  so that the times 
are compatible with an exponential distribution,  
as expected for an interior crisis  \cite{Grebogi87,Munoz2012}. 
We then fit an exponential decay to the distribution to obtain the characteristic trapping times 
$\tau_{A}$ in phase $A$ and plot them versus Reynolds number in Fig. \ref{fig_InteriorCrisis}b.
Approaching the crisis point from above, the time in phase $A$ diverges since $B$ is never visited.
According to  \cite{Grebogi87,Lai2011} the characteristic time varies as 
\begin{equation}
\tau_{A} \propto (\Rey-\Rey_{IC})^{-\gamma} \label{eqn_IC}
\end{equation}
with an exponent $\gamma$.
We use $\Rey_{IC}=785.1$, as it is the lowest values of $\Rey$ for which we observe excursions to phase $B$, 
and fit the the exponent to the data.
In the present case we obtain a good fit to the data with  $\gamma=0.8$.

The exponents for the interior crisis (as well as those of the exterior crisis) are expected to be larger or equal to $1/2$ 
for a smooth dynamical system  \cite{Grebogi1986,Lai2011}.
For one dimensional maps with a quadratic maximum they can be shown to be exactly $1/2$  but in higher dimensional systems
the folding of the manifolds contributes to the dimensions, and higher exponents have been found  \cite{Ditto1989,Tel2008,Lai2011,KreilosPRL2014}. 
Since the exponents depend on the eigenvalues at the point of bifurcation 
different crises can show different exponents, even in the same physical system.

Typical turbulent trajectories show an enormous temporal and spatial 
complexity that is difficult to create in a sequence of simple
Neimark-Sacker or period-doubling bifurcations. 
As is evident form Fig. \ref{fig_Intermittency} the dynamics of 
the system is rather regular (but not periodic!) before the interior crises and becomes increasingly
more complicated (both in the range covered and in the complexity of the time-signal) as the
Reynolds number increases.
Thus, the interior crisis bifurcation increases the complexity of the chaotic trajectories 
more dramatically than other local bifurcations and are an important contribution towards
more turbulent time evolutions. 

\begin{figure}
\centering
\includegraphics[width=0.40\textwidth]{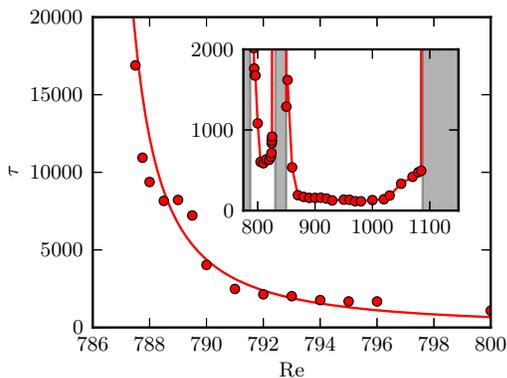}
\caption{Characteristic lifetimes $\tau$ vs. Reynolds number above the boundary crisis.
The continuous line is an algebraic fit to equation (\ref{eqn_IC}) with $\delta=1.5$.
The inset shows $\tau$ for a larger range in $\Rey$. The points are connected to guide the eye only. 
Regions where a stable attractor exists are shaded grey; the lifetime $\tau$ is infinite in these regions. }
\label{fig_LTvsRe}      
\end{figure}

\begin{figure}
\centering
\includegraphics[width=0.40\textwidth]{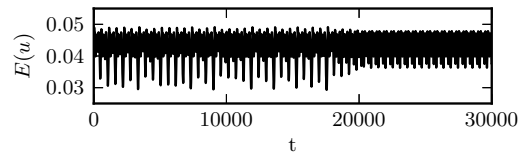}
\caption{Transition between saddle $A_{1b}$ and attractor $A_{1}$ in 
the energy trace of a trajectory at $\Rey=780.7$. 
The trajectory stays on the chaotic saddle $A_{1b}$ for about $18000$
time units before it suddenly switches to the stable attractor $A_{1}$.
\label{fig_transientA1b}}
\end{figure}

The second phenomenon we want to address here is the change in the dynamics 
near $\Rey = 786$, where the chaotic attractor suddenly disappears.
Here, the attractor collides with the lower-branch state and turns into a chaotic saddle in
a boundary or exterior crisis bifurcation  \cite{Grebogi1982}. 
It is a generic property of a chaotic saddle that the survival probabilities are exponentially distributed.
To quantify this defining property of the boundary crisis, the survival probabilities for 
$\Rey>\Rey_{XC}$ are calculated using the methods described by  \cite{Avila2010}.
The survival probabilities are clearly exponential distributed with characteristic lifetimes that depend on the
Reynolds number, as shown in Fig. \ref{fig_LTvsRe}.
As in the case of the interior crisis they diverge as
\begin{equation}
\tau \propto (\Rey-\Rey_{XC})^{-\delta} \label{eqn_XC}
\end{equation}
for $\Rey$ near the Reynolds number $\Rey_{XC}$ of the crisis bifurcation.
We fix $\Rey_{XC}=785.95$, since this is the lowest Reynolds number where
we observed a trajectory that decays after showing transient chaotic dynamics for a long time.
Best fits to the data are obtained for $\delta=1.5$, as expected for an exterior crisis.
A dense sampling of initial conditions in the state space of the system combined with
a fine scan of Reynolds numbers in the range 
between $\Rey = 778.3$ and $\Rey = 780.6$ 
reveals a small attractor $A_{1a}$ inside of $A_{1}$. 
This attractor disappears at $\Rey = 780.6$ in another boundary crisis bifurcation 
and above this Reynolds number initial conditions exist that transiently visit the saddle created by the boundary crisis of the attractor $A_{1b}$ before
suddenly settling down on $A_{1}$. 
A example for such a trajectory is shown in Fig. \ref{fig_transientA1b}. Furthermore, no initial conditions can be found that
transiently visit the saddle before becoming laminar.  This behavior is strong evidence that $A_{1a}$ 
lies completely inside of the basin of $A_{1}$.

The presence of the chaotic saddle created in the boundary crisis of $A_{1a}$ should lead to a 
second slope in the lifetime distribution as also seen in  \cite{KreilosPRL2014}.  
But since the basin of $A_{1a}$ is very small compared to $A_{1}$ this slope  
does not influence the characteristic lifetimes in Fig. \ref{fig_LTvsRe}.
The lifetimes for a larger range in \Rey are shown in the inset of Fig. \ref{fig_LTvsRe}. 
Following the boundary crisis the  lifetimes of the chaotic transients first decrease, then
start to increase again around $\Rey \approx 815$, and diverge at $\Rey = 828$,
where a second stable attractor ($A_{2}$) appears. At slightly higher \Rey 
another attractor  ($A_{3}$) appears
so that including the laminar state for a small range in $\Rey$ the system has three attracting states.
$A_{2}$ and $A_{3}$  disappear in a boundary crises at $\Rey =  837.5$ 
and $\Rey  =   841.8$, respectively.
After the boundary crisis of $A_{3}$ the lifetimes drop to even lower values than before 
the appearance of $A_{2}$. They decrease until $\Rey = 930 $, where a lifetime of $126$ 
is reached. 
Afterwards lifetimes increase again and eventually diverge at $\Rey = 1087$, 
where another attractor $A_{4}$ appears.
The attractors $A_{2}$-$A_{4}$ appear in regions of the state space occupied by the
large saddle created in the boundary crisis of $A_{1}$,
as was checked using slices of the state space as in  \cite{KreilosPRL2014}.

The crisis bifurcations analyzed here for PPF 
extend previous observations 
on Couette flow \cite{KreilosPRL2014} and also pipe flow  \cite{Altmeyer2015} in that
they provide further examples of smaller chaotic saddles 
inside larger outer saddles and 
of bifurcations that contribute to an increase of the characteristic lifetimes and 
eventually to a more complex temporal dynamics.
Moreover, the interior crises contribute to a
more dense coverage of the state space of the system by the dynamics. Thereby,
they pave the way for the transition to a chaotic saddle when the attractor collides 
with the saddle from the original saddle-node bifurcation. The connection to 
the phenomenology of low-dimensional dynamical systems and the
appearance in a number of canonical flows suggests that this transition scenario 
is typical for the transition in shear flows.

We thank John Gibson for providing {\it channelflow} and the participants and organizers of
the EUROMECH Colloquium EC565 ``Subcritical transition to turbulence'' 
(Cargese, May 2015) for fruitful  discussions. 
This work was supported by the Deutsche Forschungsgemeinschaft within FOR 1182.

\bibliographystyle{apsrev4-1}
%

\end{document}